\def\year{2022}\relax
\documentclass[letterpaper]{article} 
\usepackage{aaai22}  
\usepackage{times}  
\usepackage{helvet}  
\usepackage{courier}  
\usepackage[hyphens]{url}  
\usepackage{graphicx} 
\usepackage{multirow}
\usepackage{natbib}  
\usepackage{caption} 
\DeclareCaptionStyle{ruled}{labelfont=normalfont,labelsep=colon,strut=off} 
\usepackage{algorithm}
\usepackage{algorithmic}

\usepackage{newfloat}
\usepackage{listings}
\floatstyle{ruled}
\newfloat{listing}{tb}{lst}{}
\floatname{listing}{Listing}
\title{Can Scale-free Network Growth with Triad Formation Capture Simplicial Complex Distributions in Real Communication Networks?}
\author{
    Mayank Kejriwal\equalcontrib,
    Ke Shen\equalcontrib
}
\affiliations{
    \textsuperscript{\rm 1}Information Sciences Institute\\
    USC Viterbi School of Engineering\\
    4676 Admiralty Way 1001\\
    Marina Del Rey, California 90292\\
    
}

\usepackage{bibentry}

\begin{document}

\maketitle

\begin{abstract}
In recent years, there has been a growing recognition that higher-order structures are important features in real-world networks. A particular class of structures that has gained prominence is known as a \emph{simplicial complex}. Despite their application to complex processes such as social contagion and novel measures of centrality, not much is currently understood  about the distributional properties of these complexes in communication networks. Furthermore, it is also an open question as to whether an established growth model, such as scale-free network growth with triad formation, is sophisticated enough to capture the distributional properties of simplicial complexes. In this paper, we use empirical data on five real-world communication networks to propose a functional form for the distributions of two important simplicial complex structures. We also show that, while the scale-free network growth model with triad formation captures the \emph{form} of these distributions in networks evolved using the model, the best-fit \emph{parameters} are significantly different between the real network and its simulated equivalent. An auxiliary contribution is an empirical profile of the two simplicial complexes in these five real-world networks.  \footnote{This paper was presented as non-archival work in the Graphs and More Complex Structures for Learning and Reasoning (GCLR) workshop, co-held with AAAI 2022.}  
\end{abstract}

\section{Introduction}

Complex systems have undergone intense, interdisciplinary study in recent decades, with network science \cite{lewis2011network}, \cite{barabasi2016network} having emerged as a viable framework for understanding complexity. While early studies in network science tended to be limited to lower-order structures like dyadic links or edges \cite{seidman1983network}, \cite{milward1998measuring}, \cite{motter2005enhancing}, \cite{hagberg2008exploring} (and later, triangles), a recent and growing body of research has revealed that deep insights can be gained from the systematic study of non-simple networks, multi-layer networks \cite{mitchison1989bounds} and `higher-order' structures \cite{xu2016representing} in simple networks. 
\begin{figure}[htb]
  \centering
  \includegraphics[height=1.4in]{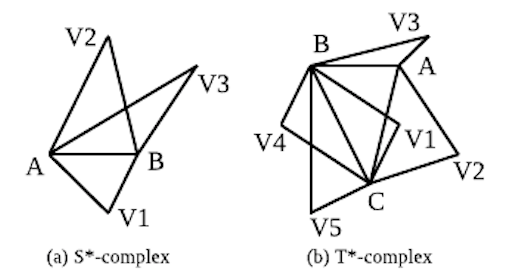}
  \caption{Illustration of an S* simplicial complex and T* simplicial complex, defined in the text.} 
  \label{fig:s_and_t}
\end{figure}

One such higher-order structure that continues to undergo study is a \emph{simplicial complex} (often just referred to as a `complex') \cite{hofmann2016complex}, \cite{barbarossa2018learning}, \cite{torres2020simplicial}. The study of simplicial complexes first took root in mathematics (especially, algebraic topology) \cite{milnor1957geometric}, \cite{faridi2002facet}, \cite{maria2014gudhi}, \cite{costa2016random}, \cite{knill2020energy}, but in the last several years, have found practical applications in network science (as discussed in \emph{Related Work}). Figure \ref{fig:s_and_t} provides a practical example of two such simplicial complexes that have been studied in the literature, especially in theoretical biology and protein interaction networks. Due to space limitations, we do not provide a full formal definition; a good reference is \cite{estrada2018centralities}, who detailed some of their properties and even proposed centrality measures due to their importance. An S-complex\footnote{Technically, we refer to these in this paper as S*- and T-* complexes, with the * indicating that we are considering the \emph{maximal} definition of the complex e.g.,  an S*-complex is not a strict sub-graph of another S-complex.} is defined  by a `central' edge A-B, with one or more triangles sharing that edge. A T-complex is similar but the central unit is a triangle (A-B-C). Furthermore, non-central (or peripheral) triangles in a T-complex should not \emph{also} participate in quads \emph{with} the central triangle i.e., given central triangle A-B-C and peripheral triangle A-B-V3, there should be no link between V3 and C in a valid T-complex. As we detail subsequently, the \emph{adjacency factor} of either an S*- or T*-complex is the number of triangles flanking the central structure (an edge or triangle respectively). 

Given the growing recognition that these two structures play an important role in real networks, and with this brief background in place, we propose to investigate the following research questions (RQs):

{\bf RQ1:} In real-world communication networks, what are the respective distributions of S*- and T*-complexes? Can good functional fits be found for these distributions?

{\bf RQ2:} Can (and to what extent) the scale-free network growth model (with triad formation) accurately capture these distributions? Or are additional parameters and steps (beyond triad formation) needed to model these higher-order structural properties in real-world networks?

\section{Related Work}
Communication networks, as well as many other natural and social networks, have the scale-free topology in common. The preferential attachment model \cite{eisenberg2003preferential}, \cite{PhysRevEpreferential} has been \emph{suggested} as a candidate network evolution or `growth' model to yield such topologies in complex networks by formalizing the intuition that highly connected nodes increase their connectivity faster than their less connected peers. The degree distribution of such networks has been shown to exhibit power–law scaling \cite{jeong2003measuring}.

While the degree distribution provides a glimpse into the structure of a complex network, models that extend pairwise relationships to multi-node relationships occurring in the system, and that allow for higher-order interactions, have been known for some time now to be important for capturing the richness and higher-order topological structures in real networks \cite{iacopini2019simplicial}, \cite{RevModPhys}, \cite{torres2020and}, \cite{BOCCALETTI2006175}, \cite{guilbeault2018complex}. In particular, in the last several years, simplicial complexes have been widely used to analyze aspects of diverse multilayer systems, including social relation \cite{wang2020social}, social contagion \cite{pastor2015epidemic}, protein interaction \cite{serrano2020simplicial}, linguistic categorization \cite{linguisticcategorization}, and transportation \cite{lin2013complex}. New measurements, such as simplicial degree \cite{serrano2020simplicial}, simplicial degree based centralities \cite{serrano2020centrality}, \cite{estrada2018centralities}, and random walks \cite{schaub2020random} have all been proposed to not only measure the relevance of a simplicial community and the quality of higher-order connections, but also the dynamical properties of simplicial networks. 

However, to the best of our knowledge, the distributional properties of such complexes, especially in the context of communication networks, have not been studied so far. A methodology for conducting such studies has also been lacking. While the former is our primary goal in this short paper, we also shed some light on the latter through our proposed methodology. 

\section{Methodology}

Since our primary goal in this paper is to understand whether (and to what extent) the scale-free network growth (with triad formation) model can accurately and empirically capture the two simplicial complexes described in the introduction, we first briefly recap the details of the growth model below. Full details are provided in \cite{triadFormation}. 

\subsection{Scale-free Network Growth with Triad Formation}
Networks with the power-law degree distribution have been classically modeled by the scale-free network model of Barab{\'a}si and Albert (BA) \cite{BAmodel}. In the original BA model, the initial condition is a network with $n_0$ nodes. In each growth timestep, an incoming node $v$ is connected using $m$ edges to existing nodes in the network. The connections are determined using \emph{preferential attachment} (PA), wherein an edge  between $v$ and another node $w$ in the network is  established with probability proportional to the degree of $w$. 


The growth model informally described above is known to generate a network with the power-law degree distribution; however, other work has found that such networks lack triadic properties (including observed clustering coefficient) in real networks. In order to incorporate such higher-order properties, the growth step in BA model was extended by \cite{triadFormation} to include a \emph{triad formation} (TF) step. Specifically, given that an an edge between nodes $v$ and $w$ was attached using preferential attachment, an edge is also established from $v$ to a random neighbor of $w$ with some probability. If all neighbors of $w$ are connected to $v$, this step does not apply.

In summary, when a `new' node $v$ comes in, a PA step will first be performed, and then a TF step will be performed with probability $P_t$ (in other words, the probability of PA without TF is $1-P_t$). These two steps are performed repeatedly per incoming node until $m$ edges are added to the network. $P_t$ is the control parameter in the model. It has been shown to have a linear relationship with the network's average (over all nodes) clustering coefficient. The clustering coefficient is a measure of the degree of \emph{clustering}, the clustering coefficient $\gamma_v$ of node $v$ is given by $\frac{|\mathcal{E}(\Gamma_v)|}{\frac{k_v(k_v-1)}{2}}$, where $|\mathcal{E}(\Gamma_v)|$ is the number of edges that exist between node $v$'s neighbors.
\begin{table}
\footnotesize
\caption{Details on five real communication networks (including average clustering coefficient) used in this paper.}

\begin{tabular}{|l|r|r|r|}
\hline
             & \multicolumn{1}{l|}{Num. Nodes} & \multicolumn{1}{l|}{Num. Edges} & \multicolumn{1}{l|}{Avg. CC.} \\ \hline
Email-Enron  & 36,265                         & 111,179                        & 0.16                            \\ \hline
Email-DNC    & 1,866                          & 4,384                          & 0.21                              \\ \hline
Email-EU     & 32,430                         & 54,397                         & 0.11                            \\ \hline
Uni. of Kiel & 57,189                         & 92,442                         & 0.04                              \\ \hline
Phone Calls  & 36,595                         & 56,853                         & 0.14                             \\ \hline
\end{tabular}

\label{table: network_details}
\end{table}
\subsection{Adjacency Factor}

To understand the distributional properties of the S*- and T*- complexes in the generated network versus real communication networks, we use the notion of the \emph{adjacency factor}. From the earlier definition, we know that an S*-complex is defined by a `central' edge (A-B in Figure 1 (a)) that is adjacent to a certain number of triangles. Given an edge in the network, therefore, we denote the adjacency factor (with respect to S*-complexes) as the (maximal) number of triangles adjacent to that edge. For example, the adjacency factor of edge A-B in Figure 1 (a) would be 3, not 1 or 2. While we record adjacency factors of 0 also\footnote{These are edges that are not part of any triangles.} to obtain a continuous distribution, only cases where adjacency factor is greater than 0 constitute valid S*-complexes.

Similarly, the adjacency factor (with respect to T*-complexes) applies to triangles in the network. For every triangle A-B-C (see Figure 1 (b)), the adjacency factor is the (maximal) number of triangles adjacent to it\footnote{But subject to the `quad' constraint noted in the \emph{Introduction}.} in the T*-complex configuration. If no (non-quad) triangles are adjacent to any of the edges of the central A-B-C triangle, then the adjacency factor is 0, meaning that the triangle does not technically participate in a T*-complex.

Hence, depending on whether we are studying and comparing S*- or T*-complex distributions, an adjacency factor can be computed for each edge and each triangle (respectively) in the network. We compute a frequency distribution over these adjacency factors to better contrast these higher-order structures in the grown versus the actual networks from a distributional standpoint.  




\section{Experiments}
\begin{figure*}[htb]
  \centering
  \includegraphics[width=6.9in]{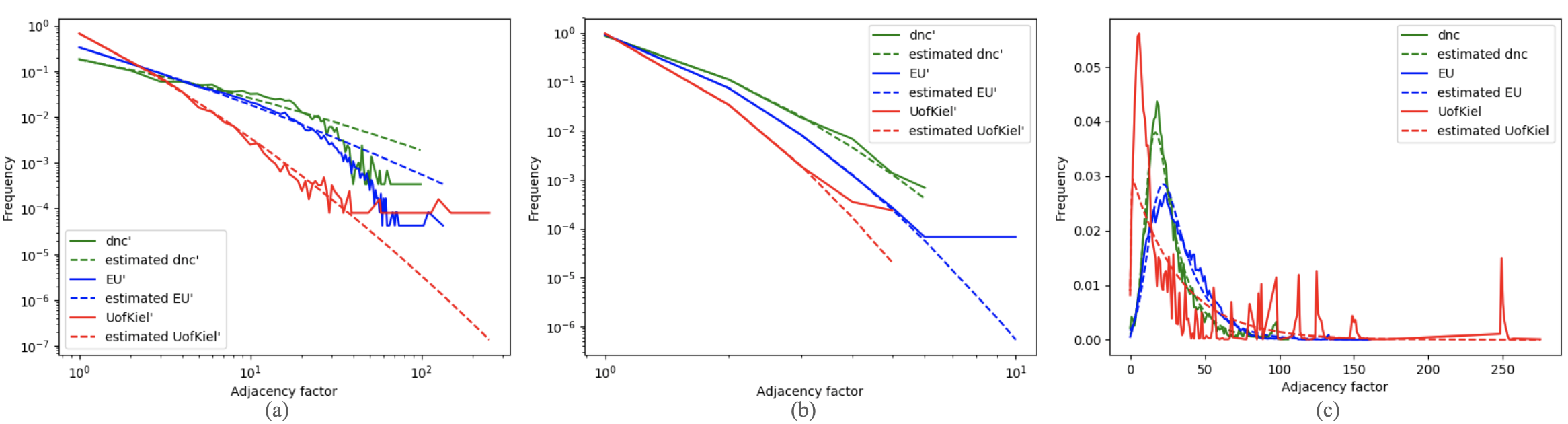}
  \caption{Frequency distribution of adjacency factors (described in the text) of (a) S*- and (c) T*-complexes in the real Email-EU, Email-DNC, and University of Kiel email networks (distributions for the other two datasets are qualitatively similar). S*-complex distributions for the generated networks are shown in (b), with the PA-TF generated networks sharing the same number of nodes, edges, and average clustering coefficient (Table 1) as their real counterparts. Due to space limitation, T*-complex frequency distribution of adjacency-factors for the generated networks is not included here. In all plots, the actual adjacency-factors distribution is always shown as a solid line and the corresponding estimated function with best-fit parameters (Equations 1 and 2) as a dashed line. Both (a) and (b) are on a log-log (with base 10) scale.} 
  \label{fig:ts-complex-distribution}
\end{figure*}
We use five publicly available communication networks in our experiments, including Enron email communication network (Email-Enron\footnote{\url{http://snap.stanford.edu/data/email-Enron.html}}), 2016 Democratic National Committee email leak network (Email-DNC\footnote{\url{http://networkrepository.com/email-dnc.php}}), a European research institution email data network (Email-EU\footnote{\url{http://networkrepository.com/email-EU.php}}), the email network based on traffic data collected for 112 days at University of Kiel, Germany \cite{uniofKiel}, and a mobile communication network \cite{phoneCall}. Details are shown in Table \ref{table: network_details}. These networks are available publicly and some (such as Enron) have been extensively studied, but to our knowledge studies involving simplicial complexes and their properties have been non-existent with respect to these communication networks. While our primary goal here is not to study these properties for these specific networks, a secondary contribution of the results that follow is that they do shed some light on the extent and distribution of such complexes in these networks. 

In the \emph{Introduction}, we had introduced two separate (but related) research questions. Below, we discuss both individually, although both rely on a shared set of results.  


{\bf RQ1:} For each network, using the numbers of nodes and edges, and the observed average clustering coefficient, we generate 10 networks using the PA-based growth model (with TF). 
We obtain the frequency distributions (normalized to resemble a probability distribution) of adjacency factors of T*- and S*-complexes in both the real and generated networks, and visualize these distributions in\footnote{The differences between generated networks corresponding to the same real network were found to be very minor, so we just show one such  network (per real network) in Figure \ref{fig:ts-complex-distribution} (b). However, subsequently described statistical analyses make use of all the generated networks. } Figure \ref{fig:ts-complex-distribution}. Besides the direct comparison between the distribution curves, the figures suggest two functions that could fit the distributions (for the S*- and T*-complexes respectively):
\begin{equation}
  \label{eqn:Equation 1 for S}
    f_{S*}(x; a, b, c) = c(bx^{-a})^{\log x}
\end{equation}
\begin{equation}
  \label{eqn:Equation 2 for T}
  f_{T*}(x;\mu, \lambda, \sigma) = \frac{\lambda}{2}e^{\frac{\lambda}{2}(2\mu+\lambda\sigma^2-2x)}erfc(\frac{\mu+\lambda\sigma^2-x}{\sqrt{2}\sigma})
\end{equation}
where $erfc(x) = \frac{2}{\sqrt{\pi}}\int_{x}^{\infty} e^{-t^2}dt$.

Both functional fits were discovered \emph{empirically} using the Enron dataset as a `development' set; however, as we show in response to RQ2, the functions fit quite consistently for all five datasets (but with different parameters, of course), although the first function diverges after a point (when the long tail begins). A theoretical basis for the functions is an interesting open question. We note that the second function is an Exponentially modified Gaussian (EMG) distribution, which is an important and general class of models for capturing skewed distributions. It has been broadly studied in mathematics, and has found empirical applications as well \cite{foley1984review}. 

\begin{table*}
\resizebox{\textwidth}{16mm}{
\begin{tabular}{|l|c|c|c|c|c|c|c|c|}
\hline
\multirow{3}{*}{} & \multicolumn{4}{c|}{Equation 1}                                 & \multicolumn{4}{c|}{Equation 2}                                          \\ \cline{2-9} 
                  & \multicolumn{4}{c|}{\textit{real / grown}}                               & \multicolumn{4}{c|}{\textit{real / grown}}                                        \\ \cline{2-9} 
                  & $a$ / $\overline{a'}$      & $b$ / $\overline{b'}$      & $c$ / $\overline{c'}$      & $MND$ / $\overline{MND'}$ / \textit{ref.} $MND$ & $\lambda$ / $\overline{\lambda'}$ & $\mu$ / $\overline{\mu'}$     & $\sigma$ / $\overline{\sigma'}$ & $MND$ / $\overline{MND'}$ / \textit{ref.} $MND$ \\ \hline
Email-Enron       & 0.25 / 0.82 & 0.75 / 0.59 & 0.19 / 0.53 & 0.68 / 0.79 / 0.71    & 0.02 / 0.21      & 6.82 / 0.87  & 5.08 / 0.79    & 0.37 / 0.83 / 0.91    \\ \hline
Email-DNC         & 0.07 / 1.25 & 0.50 / 0.15 & 0.18 / 0.85 & 0.64 / 0.33 / 0.71    & 0.07 / 0.76      & 10.86 / 0.00 & 5.06 / 0.00    & 0.33 / 0.53 / 0.90    \\ \hline
Email-EU          & 0.06 / 1.79 & 0.33 / 0.09 & 0.33 / 0.92 & 0.67 / 0.34 / 0.74    & 0.05 / 1.59      & 13.81 / 0.00 & 7.58 / 0.00    & 0.70 / 0.84 / 0.90    \\ \hline
Uni. of Kiel      & 0.16 / 1.35 & 0.15 / 0.02 & 0.67 / 0.97 & 0.49 / 0.47 / 0.55    & 0.03 / 3.13      & 0.37 / 0.00  & 0.57 / 0.00    & 0.68 / 1.07 / 0.74    \\ \hline
Phone Calls       & 0.65 / 1.78 & 0.44 / 0.11 & 0.55 / 0.90 & 0.56 / 0.28 / 0.84    & 0.43 / 1.42      & 0.00 / 0.00  & 0.00 / 0.00    & 0.44 / 0.79 / 0.76    \\ \hline
\end{tabular}}
\caption{Best-fit parameter estimates for both Equations 1 and 2. The MND and reference baseline is described in the text.}
\label{table: fit_para}
\end{table*}
{\bf RQ2: } We tabulate the best-fit parameters for each real world network, and the generated networks, in Table \ref{table: fit_para}. For the real world networks, there is only one set of best-fit estimates. For the generated networks there are ten best-fit estimates per parameter (since we generate 10 networks per real-world network), for which we report the average in the table. We also compute a 2-tailed Student's t-test and found that, for all parameters and all networks, the generated networks' (averaged) parameter is significantly different from the corresponding real network's best-fit parameter at the 99\% level. This suggests, intriguingly, that despite the similarities between distributions in Figure \ref{fig:ts-complex-distribution} (a) and (b)  (i.e., between the real and generated networks) the best-fit parameters (for real networks) are significantly different in both cases.

Of course, this does not answer the question as to whether the functions that we empirically discovered and suggested in Equations 1 and 2 are good approximations or models for the actual distributions. To quantify such a `goodness of fit' between an actual frequency distribution curve (whether for the real network or the generated networks) and the curve obtained by using the models suggested in  Equation 1 or 2 (with best-fit parameter estimates), we compute a metric called \emph{Mean Normalized Deviation} (MND). This metric is modeled closely after the root mean square error (RMSE) metric. Given an actual curve $f$ and a modeled curve $f'$, defined on a common support\footnote{In our case, this is simply the set of adjacency factors.} (x-axis) $X=\{x_1, x_2, \ldots x_n\}$, the MND is given by:

\begin{equation}\footnotesize
   MND(f,f', X) = \frac{1}{|X|} \sum_{x \in X} \frac{|f'(x)-f(x)|}{f(x)} 
\end{equation}

Note that the lower the MND, the better $f'$ fits $f$ on support $X$. The MND can never be negative for a positive function $f$, but it has no upper bound. Hence, a reference is needed. Since we are not aware of any other candidate functional fits for the simplicial complex distributions in the literature, we use a simple (but functionally effective) baseline, namely the horizontal curve $y=c$, where $c$ is a constant that is selected to roughly coincide with the long-tail of the corresponding real network's distribution. 

In Table 2, we report not just the MNDs of the real and grown networks, but also the corresponding reference MND. Because of the significant long tails in Figure 2, this MND is already expected to be low. We find, however, that with only three exceptions (over both equations\footnote{Specifically, on the Equation 1 model, the MND' (average over generated networks) is higher for \emph{Email-Enron} than for the reference; on Equation 2, the MND' is higher for both \emph{Uni. of Kiel} and \emph{Phone Calls}.}) for the generated networks (and none for the real networks) does the reference fit the actual distributions better than our proposed models (through a lower MND), despite being optimized to almost coincide with the long tail. Interestingly, Equation 2 has (much) lower MND scores for the real network compared to the grown networks, as well as the reference function. As we noted earlier, Equation 1 did not seem to be capturing the long tail accurately. We hypothesize that a piecewise function, where Equation 1 is only used for modeling the short tail of the S*-complex frequency distribution, may be a better fit. In all cases, investigating the theory of this phenomenon is a promising area of investigation for complex systems research.

\section{Conclusion}

Simplicial complexes have become important in the last several years for modeling and reasoning about higher-order structures in real networks. Many questions remain about these structures, including whether they are captured properly by existing (and now classic) growth models. In this paper, we showed that, for two well-known complexes, the PA-model with triad formation captures the distributional properties of the complexes, but the best-fit parameters are significantly different between the grown networks and the real communication networks. It remains an active area of research to better understand the theoretical underpinnings of our proposed functional fits 
for the simplicial complex distributions, and also to deduce what could be `added' to the growth model to bring its parameters into alignment with the real-world network. We are also investigating the properties of other growth models with respect to accurately capturing these distributions. Finally, understanding the real-world phenomena modeled by these complexes, which are fairly common motifs in all five networks we studied, continues to be an interesting research avenue in communication (and other complex) systems. 
\newpage
\bibliography{aaai22}

\end{document}